\documentclass[twocolumn,aps,prl,floatfix,superscriptaddress]{revtex4-1}
\usepackage{amsmath,amssymb}
\usepackage{graphicx,float}
\usepackage{times,txfonts}
\usepackage{color}
\usepackage{multirow}
\usepackage{bbold}
\usepackage{color}
\usepackage{soul}
\usepackage{hyperref}
\usepackage{times,txfonts}
\usepackage{natbib}
\usepackage{blindtext}
\usepackage[export]{adjustbox}
\usepackage{mathrsfs}
\usepackage{textcomp}
\usepackage{blkarray}
\usepackage{multirow}
\newcommand{\bra}[1]{\left\langle #1 \right|}
\newcommand{\ket}[1]{\left| #1 \right\rangle}
\newcommand{\braket}[2]{\left\langle {#1{\left| \vphantom{#1 #2} \right.} #2} \right\rangle}

\renewcommand{\epsilon}{\varepsilon}

\def\VR{\kern-\arraycolsep\strut\vrule &\kern-\arraycolsep}
\def\vr{\kern-\arraycolsep & \kern-\arraycolsep}
\usepackage{sistyle}
\usepackage{soul}
\definecolor{lightblue}{RGB}{185,210,248}
\sethlcolor{lightblue}

\begin{document}
\title{Round-Robin Differential Phase-Shift Quantum Key Distribution with Twisted Photons}
\author{Fr\'ed\'eric Bouchard}
\email{fbouc052@uottawa.ca}
\affiliation{Department of Physics, University of Ottawa, Advanced Research Complex, 25 Templeton Street, Ottawa ON Canada, K1N 6N5}
%%%
\author{Alicia Sit}
\affiliation{Department of Physics, University of Ottawa, Advanced Research Complex, 25 Templeton Street, Ottawa ON Canada, K1N 6N5}
%%%
\author{Khabat Heshami}
\affiliation{National Research Council of Canada, 100 Sussex Drive, Ottawa ON Canada, K1A 0R6}
%%%%
\author{Robert Fickler}
\affiliation{Department of Physics, University of Ottawa, Advanced Research Complex, 25 Templeton Street, Ottawa ON Canada, K1N 6N5}
%%%%
\author{Ebrahim Karimi}
\affiliation{Department of Physics, University of Ottawa, Advanced Research Complex, 25 Templeton Street, Ottawa ON Canada, K1N 6N5}
\affiliation{Department of Physics, Institute for Advanced Studies in Basic Sciences, 45137-66731 Zanjan, Iran.}
%%%%%%%%
%
%\pacs{Valid PACS appear here}% PACS, the Physics and Astronomy

%Nature Photonics, Letter: 150 words abstract, main text of no more than 1,500 words and 5 display items (figure, tables); references are limited to 30. Section headings are not used.
%Nature Photonics, Article: 150 words abstract, main text of no more than 3,000 words and 6 display items (figure, tables); references are limited to 30. Section headings are not used.

%PRL: 3750 words for main text 

\begin{abstract}
Quantum key distribution (QKD) offers the possibility for two individuals to communicate a securely encrypted message. From the time of its inception in 1984 by Bennett and Brassard, QKD has been the result of intense research. One technical challenge is the monitoring of signal disturbance in a QKD system to bound the information leakage towards an unwanted eavesdropper. Recently, the round-robin differential phase-shift (RRDPS) protocol, which encodes bits of information in a high-dimensional state space, was proposed to solve this exact problem. Since its introduction, many realizations of the RRDPS protocol were demonstrated using trains of coherent pulses. Here, we propose and experimentally demonstrate an implementation of the RRDPS protocol using the photonic orbital angular momentum degree of freedom. In particular, we show that Alice's generation stage and Bob's detection stage can each be reduced to a single phase element, greatly simplifying its implementation. Our scheme offers a practical demonstration of the RRDPS protocol which will suppress the need for monitoring signal disturbance in free-space channels.
\end{abstract}
%166

\maketitle

%Intro to QKD
The early protocols of QKD, such as the BB84 protocol and others~\cite{bennett:84,ekert:91,bennett:92,gisin:02}, demonstrated the power of quantum cryptography using simple and elegant schemes. However, it has become increasingly evident that their implementations lead to unforeseen challenges and technical difficulties~\cite{scarani:09}. This motivates more practical protocols that are designed towards specific implementations, and also provides theoretical tools for a complete security analysis of a QKD system. In general, this leads to less elegant and more complicated schemes. An important example of such a scheme aimed towards practicality is the \emph{plug-and-play setup}~\cite{muller:97}. A major limiting factor in QKD implementations is the photon source. Many practical implementations rely on weak coherent states, which have a non-zero probability of generating multi-photon events. Surprisingly, a slight modification of the BB84 protocol in the announcements (SARG04)~\cite{scarani:04} leads to an improvement against \emph{photon-number-splitting} (PNS) attacks~\cite{brassard:00}. A further improvement may be achieved by using the \emph{decoy states} protocol~\cite{hwang:03}. An example of a protocol that incorporates these different strategies is the \emph{coherent one way} (COW) protocol~\cite{stucki:05}, which can operate at high speeds, and can be integrated into existing fibre networks.
%180

Another major challenge in any QKD system is the treatment of noise in the quantum channel. To ensure security, it is necessary to assume the worst case scenario: noises are the result of an adversary, referred to as \emph{Eve}, eavesdropping on the channel. The authorized partners, \emph{Alice} and \emph{Bob}, may use \emph{error correction} protocols to remove any errors in their shared key. Any information leakage to Eve may be subsequently removed using \emph{privacy amplficiation}~\cite{bennett:95}. However, in the presence of large environmental noises, users must abandon their link if the errors are above a given threshold imposed by the security of the QKD protocol. A promising avenue to increasing error tolerability in noisy environments is the use of high-dimensional systems, known as \emph{qudits}, rather than \emph{qubits}~\cite{bechmann:00a,bechmann:00b,cerf:02}. Additionally, high-dimensional QKD protocols are shown to transmit more information per carrier. A natural extension of the BB84 protocol leads to a secret key rate, in the infinite-sized-key limit, of $R=\log_2(d)-2 h^{(d)}(e_b)$, where $d$ is the dimension, $e_b$ is the QBER and $h^{(d)}(\cdot)$ is the $d$-dimensional Shannon entropy.%169

%%%%%%%%%%%%%%
\begin{figure*}[t]
	\begin{center}
	\includegraphics[width=1.8\columnwidth]{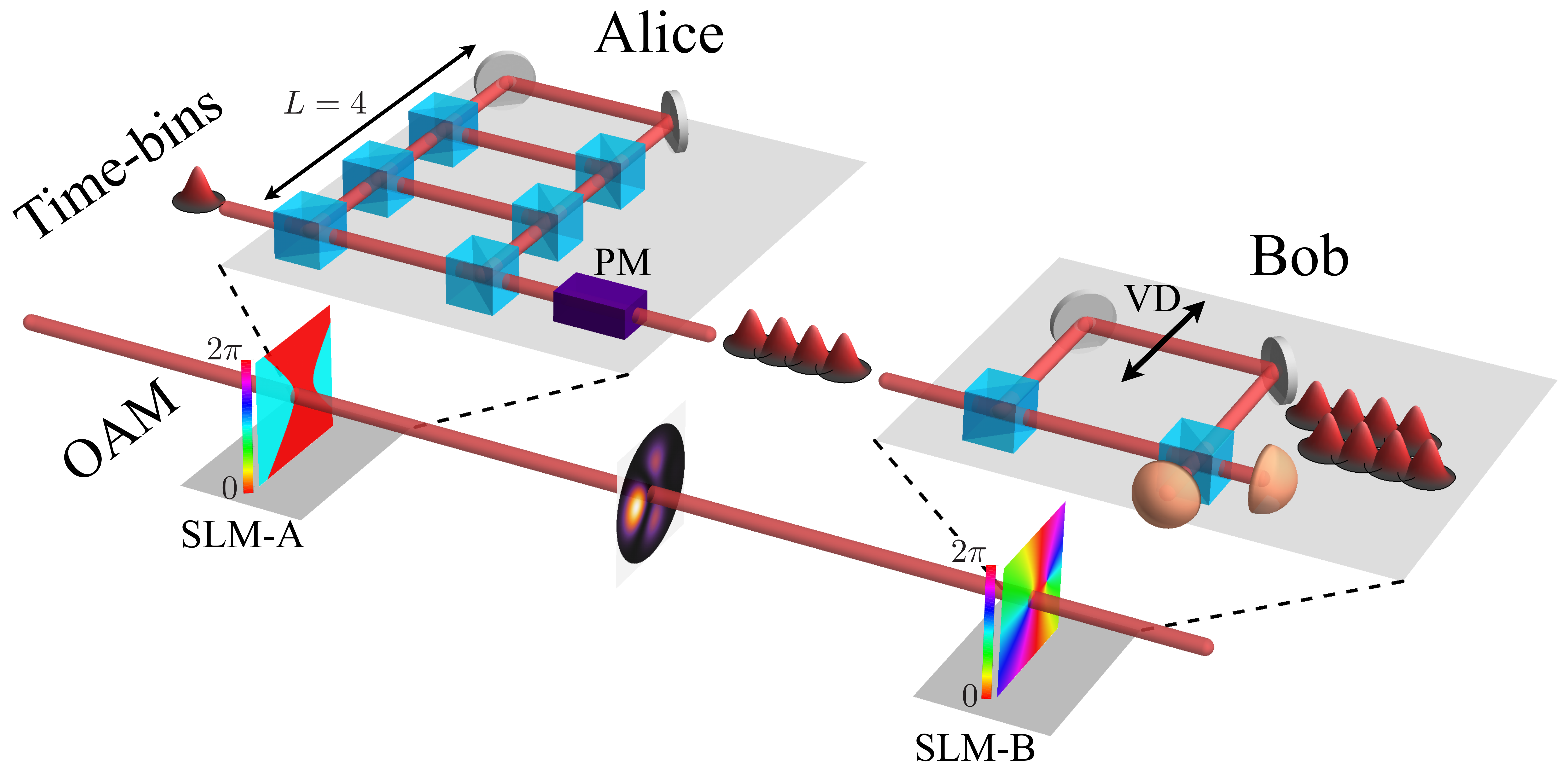}
	\caption[]{Experimental setup: The original RRDPS scheme using the phase of time-bins to encode information (depicted in the back) consisting of multiple interferometers for generation with a phase modulator (PM), and another interferometer with a variable delay (VD) for detection. By encoding using OAM instead, we can reduce the setups to be two phase holograms, displayed on spatial light modulators (SLMs), one for generation and one for detection of the states. An example of Alice's generation phase element for a state with $L=4$, i.e. $s_\ell=(0,0,0,0)$ is shown on her SLM-A. The intensity distribution of the generated state is shown in the channel, followed by Bob's measurement phase element (SLM-B) with a shift of $r=1$ and a detected OAM value of $m=2$ for a relative phase of 0.}%120
	\label{fig:setup}
	\end{center}
\end{figure*}
%%%%%%%%%%%%%%

Another class of high-dimensional QKD protocols are the \emph{distributed-phase-reference} (DPR) schemes which, unlike the high-dimensional BB84 protocol, encode only one bit of information per carrier. This reduction in transmitted information comes with the benefit of increased error tolerance. The original differential phase shift (DPS) protocol~\cite{inoue:02} employs superpositions of high-dimensional states where the information is encoded in the relative phase of the qudits. In the differential phase \emph{Chau15} protocol~\cite{chau:15,chau:17}, the information is encoded in the relative phase of a ``qubit-like'' high-dimensional state~\cite{wang:17}. The Chau15 protocol has the particular advantage of tolerating error rates of up to 50~\%. It is also possible to achieve a DPR scheme where more than one bit of information is transmitted~\cite{bacco:16}. Recently, a QKD scheme has been proposed in which the amount of information leakage is bounded by Alice's state preparation stage and Bob's measurement unit, and therefore eliminates the requirement for monitoring the signal disturbance~\cite{sasaki:14}. This scheme, known as \emph{round-robin differential phase-shift} (RRDPS) QKD, is based on encoding a random bit sequence of length $L$ in the phase of a coherent $L$-dimensional superposition state. After transmission through a quantum channel accessible to Eve, Bob randomly selects the setting of his interferometric measurement to record one classical bit from a phase difference measurement between two modes of the $L$-dimensional input state. In this Letter, we propose and experimentally demonstrate an implementation of RRDPS QKD based on the transverse mode of photons, in particular using the orbital angular momentum (OAM) degree of freedom. Our proposed scheme simplifies the experimental setup for generation and detection with the ability to flexibly implement different dimensions, which we demonstrate for $L=3$ to 8, along with 16, 32 and 64. 
%270

The original proposal for implementation of the RRDPS QKD is based on state preparation in an $L$-pulse train single-photon state, and phase difference measurement using a switchable delayed interferometer with variable delays, see Fig.~\ref{fig:setup}. Several experimental demonstrations of RRDPS QKD have used this encoding with passive~\cite{guan:15} or active variable delays~\cite{takesue:15,wang:15,li:16} using time-bin state dimensions of $L=5,~65$ and $128$. The secret key rate, in the infinite-sized-key limit, is given by \mbox{$R=1-h^{(2)}(e_b)-h(1/(L-1))$}~\cite{sasaki:14} and was recently improved to \mbox{$R=1-h^{(2)}(e_b)-\mathrm{max}_{0\leq x \leq 1}  \varphi [ (L-1)x,1-x]/(L-1)$}~\cite{Yin:18}, where \mbox{$\varphi[x,y]=-x \log_2 x -y \log_2 y +(x+y) \log_2 (x+y)$}. Although these experiments highlight the described benefits in terms of error thresholds, they also demonstrate the expected challenges in implementing Bob's measurement unit that requires switchable delays and active stabilization of delayed interferometers. Given the practical implications of this scheme, it is desirable to explore encoding in other degrees of freedom of photons in search of simplified implementations. Here, we investigate encoding on the OAM degree of freedom.
%144

%Talk about high-dimensional QKD with OAM
Photons carrying OAM are characterized by a exp($i\ell\phi$) phase factor~\cite{allen:92}, where $\ell$ is an integer and $\phi$ is the transverse azimuthal angle, giving rise to $\ell$ intertwined helical wavefronts and a null of intensity along the propogation axis. The OAM of photons provides an unbounded Hilbert space~\cite{erhard:17}, limited in practice by the numerical aperture of the system. OAM has provided a simplified way of realizing high-dimensional protocols. Only a single phase element is required for generation --- typically a hologram displayed on a programmable spatial light modulator (SLM)~\cite{heckenberg:92,forbes:16}. For detection, a second SLM and single mode fibre (SMF)~\cite{mair:01,qassim:14} or OAM sorter~\cite{berkhout:10,larocque:17} can be implemented to efficiently filter the different modes. High-dimensional QKD protocols using OAM have been successfully demonstrated in the laboratory~\cite{mafu:13,mirhosseini:15}, studied and realized in realistic free-space conditions~\cite{vallone:14,krenn:15,sit:17}, and recently implemented in an underwater quantum channel~\cite{bouchard:18}. Using these encryption techniques with OAM, we adapt previous RRDPS QKD schemes, which use many interferometers, to consist of two phase elements --- one for state generation by Alice, and one for state measurement by Bob, shown in Fig.~\ref{fig:setup}. In the original DPS proposal, Alice's generation is done by sending a single photon into an interferometer consisting of $L$ arms. However, for practical considerations, a sequence of intensity modulators, phase modulators and attenuator accomplishes the desired generation stage of Alice using weak coherent pulses. It is Bob's detection stage that presents the main technical challenges. Though using OAM modes of light is not as scalable as time-bins, their generation and detection techniques do not require active stabilization. 
%248

%%%%%%%%%%%%%%
\begin{figure*}[!htbp]
	\begin{center}
	\includegraphics[width=1.8\columnwidth]{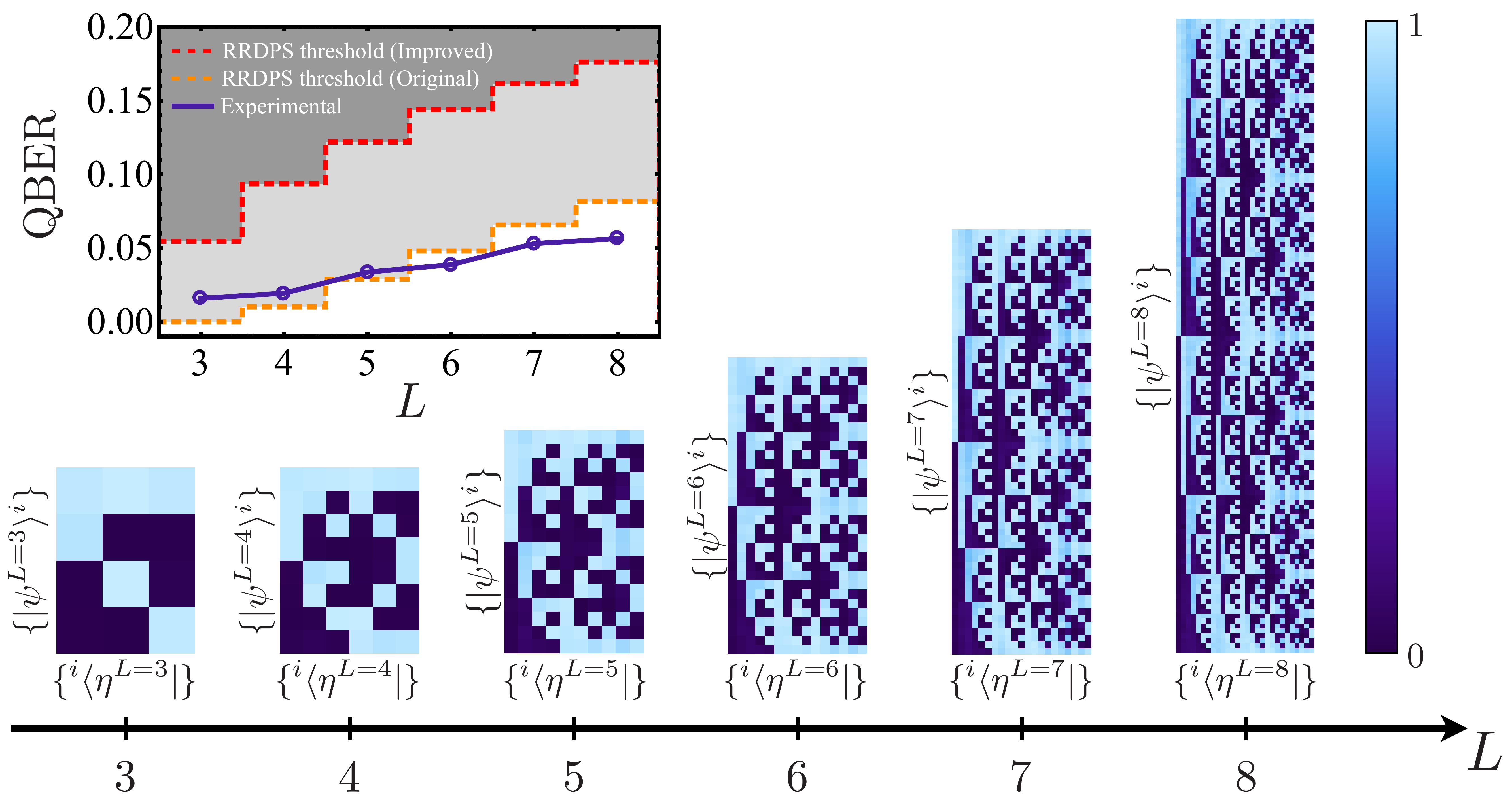}
	\caption[]{Experimental Results: Measured probability-of-detection matrices for RRDPS QKD scheme with OAM for dimensions $L=3$ to $L=8$. The rows consist of the $2^{L-1}$ states, $\left\{ \ket{\psi}^i \right\}$, that Alice generates. The columns are the $L(L-1)/2$ measurements, $\left\{ ^i \bra{\eta} \right\}$, that Bob performs to determine the relative phase between two of the OAM states. Here, we are showing the results for when the two states are in-phase, i.e. relative phase of 0. The inset graph compares the quantum bit error rate (QBER) with increasing dimension between the theoretical original~\cite{sasaki:14} and improved RRDPS~\cite{Yin:18} error thresholds and our experimentally measured QBER.}%90
	\label{fig:fig2}
	\end{center}
\end{figure*}
%%%%%%%%%%%%%%

Let us start with the preparation stage: Alice prepares an input state by encoding a random bit sequence of length $L$ in the phases of different OAM states of a single photon,
	\begin{equation}\label{input_state}
		\ket{\psi^L} = \frac{1}{\sqrt{L}}\sum_{\ell}(-1)^{s_{\ell}}\ket{\ell},
	\end{equation}
where $s_{\ell} \in \{ 0,1 \}$, $\ket{\ell}$ are the OAM eigenstates that represent a single photon carrying $\ell\hbar$ units of OAM, analogous to the photon's time bin in the pulse train of the original RRDPS protocol, where $\hbar$ is the reduced Planck constant. For $L$ even and odd, we respectively consider \mbox{$\ell \in \{-L/2, ... , L/2 ; \ell \neq 0 \}$} and \mbox{$\ell \in \{ -(L-1)/2, ...,  (L-1)/2 \}$}. Alice can generate this state with a single computer-generated hologram displayed on her SLM. One of the main advantage of using OAM states of photons is the inherent stability and interference visibility of superposition states. Superpositions of OAM states simply correspond to another transverse spatial mode of light, which in practice preserve the same stability as OAM modes. In spite of the fact that free-space propagation may alter the OAM superposition state due to Gouy phases~\cite{born:13}, this effect is straightforward to control and compensate for. As one considers larger values of $L$, the stability of the OAM state does not degrade as it might be in the case of an interferometric configuration. However, the scalability is limited by the numerical aperture of the system.
%205

Alice's photon is sent through the quantum channel and received at Bob's detection setup. In order to extract one qubit of information, following the original protocol, Bob sends the state into an interferometer and \emph{shifts} one arm by a randomly chosen amount \mbox{$r\in\{1,L-1\}$}. For time-bins encoding, the shift corresponds to delaying each pulse in the train by $rT$, where $T$ is the time between pulses. This \emph{shift}-type interferometer may be achieved with OAM by inserting a phase element in one arm of a balanced interferometer. The phase element is given by $\exp (i r \phi)$, where this has the effect of shifting the OAM value analogously, i.e. $\ket{\ell} \longrightarrow \ket{\ell+r}$. However, such configuration does not take full advantage of the OAM scheme. It can be seen that an interferometer with a phase shift element in one arm and no phase element in the other arm is equivalent to a single phase element given by $(1 \pm \exp (i r \phi))/\sqrt{2}$, where the relative phase ($\pm$) determines the ``output port" of the interferometer. The last step in the RRDPS detection scheme is the time-resolved detection of the interfered pulse train. This is equivalent to any OAM detection scheme such as mode filtering~\cite{mair:01} or OAM sorting~\cite{berkhout:10}. Mode filtering is achieved using a phase element given by $\exp(- i m \phi)$, which corresponds to a projection onto the state $\ket{m}$, followed by coupling into a SMF. A further simplification is done by combining the ``measurement" phase element with the ``shifting-interferometer" phase element, resulting in a phase of $\exp(- i m \phi) ( 1\pm \exp (i r \phi))/\sqrt{2}$. This can be seen as a projection onto the state $\ket{r,m} = (\ket{m} \pm \ket{m-r} )/\sqrt{2}$, labelled by the random shift of $r$ and a projection onto the state with OAM value of $m$, i.e. $\left| \braket{\psi^L}{r,m} \right|^2$. Our simplified detection configuration corresponds to a filter-based scheme, resulting in a lower detection efficiency. However, by replacing the phase-flattening component with a sorter-type measurement device, the pre-sifting detection efficiency can reach unity. 
%307

In the event of detecting a click from displaying $(\ket{m}- \ket{m-r} )/\sqrt{2}$ or $(\ket{m}+\ket{m-r} )/\sqrt{2}$, Bob records a 0 or 1, respectively. In order for Alice to have a shared corresponding bit, Bob publicly announces the random indices $m$ and $r$. Alice obtains her sifted key bit by computing $s=s_m \oplus s_{m-r}$, where $\oplus$ corresponds to summation modulo 2. Moreover, Alice and Bob only keep the outcomes where $m \leq L$ with a sifting efficiency of 1/2, as in the case of BB84.
%72

Based on the proposed simplified scheme above, we perform a proof-of-principle experimental realization of the RRDPS protocol using OAM encoded on heralded single photons. We pump a 3~mm $\beta$-barium borate type I nonlinear crystal with a quasi-continuous wave UV laser at 355~nm to produce single photon pairs (signal and idler) at 710~nm via spontaneous parametric down-conversion. The photons are spatially filtered to the fundamental Gaussian mode by coupling them to SMF with a measured single photon heralding coincidence rate of 40~kHz within a coincidence window of 5~ns. Alice prepares the signal photon into an equally weighted superposition of OAM states with a phase-only SLM, SLM-A, where a string of $L$ bits is encoded in the phase (0 or $\pi$) of each OAM state. She subsequently distributes the photon to Bob through one meter of free-space in the laboratory. We note that when considering all $2^L$ possible bit sequences, the second half of the corresponding states are identical to the first half with an additional global phase of $\pi$. Hence, Alice randomly chooses one state out of the set of $2^{L-1}$ possible superposition states, $\left\{ \ket{\psi^L}^i \right\}$, with distinct bit sequences. Bob projects the received state, using his own SLM (SLM-B) and SMF, onto a particular superposition of two OAM states from a choice of $L(L-1)/2$ possibilities, i.e. $\left\{ \ket{\eta} = \ket{r,m} \right\}$. The idler photon is used to herald the presence of the signal photon for a coincidence event after Bob's measurement.
%237

%%%%%%%%%%%%%%
\begin{figure}
	\begin{center}
	\includegraphics[width=\columnwidth]{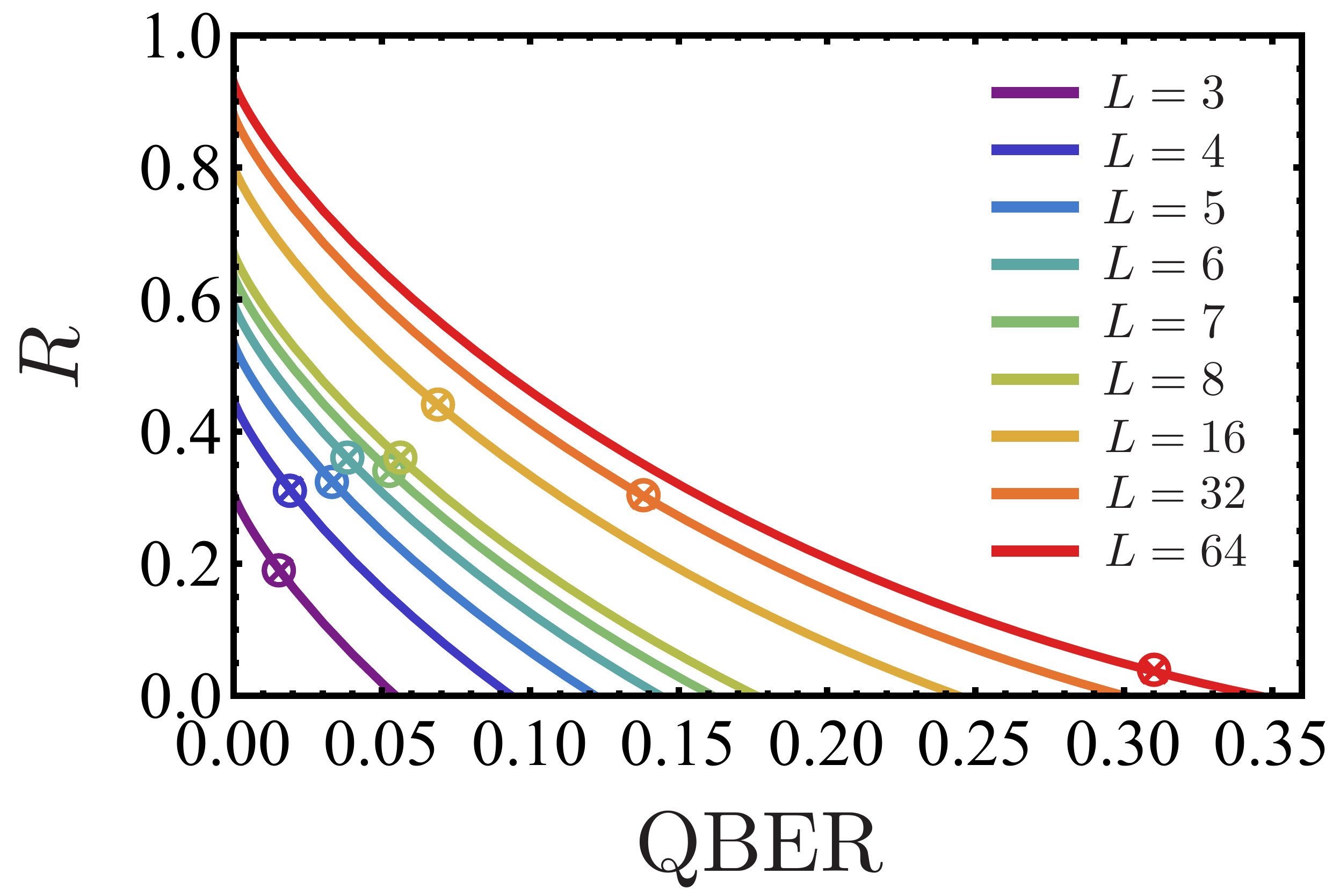}
	\caption[]{Secret key rates: Calculated secret key rates (circles), $R$, for dimensions $L=3$ to 8, 16, 32 and 64 from the experimentally measured QBERs with respect to the improved theoretical bounds (solid curves). As the dimension increases, the number of transmitted bits per sifted photon increases until we are limited by the numerical aperture of our system for higher dimensions.}%58
	\label{fig:fig3}
	\end{center}
\end{figure}
%%%%%%%%%%%%%%

%%%%%%%%%%%
%\begin{table}
%\centering
%	\begin{tabular}{lccc}
%	\hline \hline
%	$L$ & QBER & $R$ (Originial) & $R$ (Improved)\\
%	\hline 
%	3 & 0.016 & - & 0.188  \\
%	4 & 0.019 & - & 0.310 \\
%	5 & 0.034 & - & 0.322 \\
%	6 & 0.039 & 0.042 & 0.358 \\
%	7 & 0.053 & 0.051 & 0.339 \\
%	8 & 0.056 & 0.095 & 0.359 \\
%	16 & 0.069 & 0.284 & 0.440 \\
%	32 & 0.139 & 0.214 & 0.301 \\
%	64 & 0.315 & - & 0.032\\
%	 \hline \hline
%	\end{tabular}
%\caption[]{Table...}
%\label{table:1}
%\end{table}
%%%%%%%%

The experiment is performed, under identical laboratory conditions, for values of $L$ ranging from 3 to 8, 16, 32, and 64; the results for 3 to 8 are shown in Fig.~\ref{fig:fig2}. The measured probability-of-detection matrices shown represent the likelihood that Bob detects a click following the projection of the incoming photon's state onto $\ket{\eta}=(\ket{m}+\ket{m-r} )/\sqrt{2}$. Figure~\ref{fig:fig2} also shows the experimentally extracted QBER with respect to dimension ($L$). Both the original~\cite{sasaki:14} and improved~\cite{Yin:18} theoretical QBER thresholds of the RRDPS protocol security analyses are also shown along with the experimental QBER. For small dimensions, i.e. $L=3$ to 5, the QBER exceeds the original error thresholds. In particular, for the case of $L=3$, the original protocol cannot generate secret key bits. However, the improved theoretical bounds show that secure communication would still be feasible in all measured dimensions. Interestingly, above $L=5$, the experimental QBERs are lower than the original thresholds, which appears to increase faster than our systematic errors do with dimension. The associated secret key rates, $R$, calculated from the improved RRDPS protocol, are given in Fig.~\ref{fig:fig3} along with the theoretical secret key rates. Values of $e_b^{L=3}=0.016$, $e_b^{L=4}=0.019$, $e_b^{L=5}=0.034$, $e_b^{L=6}=0.039$, $e_b^{L=7}=0.053$, $e_b^{L=8}=0.056$, $e_b^{L=16}=0.069$, $e_b^{L=32}=0.139$, and $e_b^{L=64}=0.315$ where obtaining experimentally for the QBER. The corresponding secret key rates are $R^{L=3}=0.188$, $R^{L=4}=0.310$, $R^{L=5}=0.322$, $R^{L=6}=0.358$, $R^{L=7}=0.339$, $R^{L=8}=0.359$, $R^{L=16}=0.440$, $R^{L=32}=0.301$, and $R^{L=64}=0.032$ bits per sifted photon, respectively.%225

For the cases of $L=16$, 32, and 64, in the interest of time, a subset of 1500 randomly selected generation and measurement settings is considered out of the $2^{L-1}L(L-1)/2$ possibilities. Interestingly, it is for the case of $L=16$ that the largest secret key rate is achieved. We see an increase error rate (decrease in secret key rate) for higher dimensions most likely due to the limiting numerical aperture of our system. One potential way to circumvent this issue would be to extend our scheme to include radial modes~\cite{karimi:14b,karimi:14c} to make use of the full transverse spatial structure, i.e. optimize the best number of modes to fill the numerical aperture.
%105

%
% Conclusion
%

Finally, another advantage of our proposed scheme is the versatility of the experimental configuration. In the first place, the phase elements may be achieved using various techniques according to practical considerations, e.g. liquid crystal devices~\cite{marrucci:06,heckenberg:92}, digital micro-mirror devices~\cite{mirhosseini:13b}, refractive elements~\cite{beijersbergen:94}, metasurfaces~\cite{yu:11,karimi:14,bouchard:14}, etc. More importantly, the experimental configuration of our proposed scheme is compatible with many other prepare-and-measure high-dimensional QKD protocols such as BB84~\cite{bennett:84}, mutually unbiased bases (MUB)-based protocols~\cite{durt:10,mafu:13}, Singapore~\cite{englert:04b}, DPS~\cite{inoue:02}, and Chau15~\cite{chau:15,chau:17}. For fluctuating environmental conditions, Alice and Bob may adapt by selecting the most favourable QKD protocol. Hence, a single experimental configuration compatible with many protocols exhibit the versatility of QKD schemes based on OAM encoding~\cite{bouchard:18b}. The RRDPS scheme is one such example where the difficulty of monitoring signal disturbance is removed and may lead to improved performances in noisy environments. It will be important to see how this scheme performs in real-world conditions, in particular where there is atmospheric turbulence. 
%174

%Text:180+169+120+270+144+248+90+205+307+72+237+58+225+105+174 = 2604
%Eq: 4 x 16 = 64
%Fig: (300/(0.5 x (1683/832))+40) + (300/(0.5 x (1536/817))+40) + (150/(0.5 x (718/480))+20) = 916
%Total: 3584 < 3750

\vspace{0.5 EM}

\noindent\textbf{Acknowledgments} All authors would like to thank Duncan England for fruitful discussions. F.B. acknowledges the financial support of the Vanier graduate scholarship of the NSERC. R.F. acknowledges the financial support of the Banting postdoctoral fellowship of the NSERC. This work was supported by Canada Research Chairs (CRC); Canada Foundation for Innovation (CFI); Canada First Research Excellence Fund (CFREF); Natural Sciences and Engineering Research Council of Canada (NSERC).

%Bibliography
\bibliographystyle{naturemag}
%\bibliography{rrdps}

\begin{thebibliography}{10}
\expandafter\ifx\csname url\endcsname\relax
  \def\url#1{\texttt{#1}}\fi
\expandafter\ifx\csname urlprefix\endcsname\relax\def\urlprefix{URL }\fi
\providecommand{\bibinfo}[2]{#2}
\providecommand{\eprint}[2][]{\url{#2}}

\bibitem{bennett:84}
\bibinfo{author}{C.~H. Bennett} and \bibinfo{author}{G. Brassard,}
\newblock \bibinfo{title}{in \textit{Proceedings of the IEEE International Conference on
  Computers, Systems, and Signal Processing} (IEEE, New York, 1984), pp.175-179.}

\bibitem{ekert:91}
\bibinfo{author}{A.~K. Ekert,}
\newblock \emph{\bibinfo{journal}{Phys. Rev. Lett.}}
  \textbf{\bibinfo{volume}{67}}, \bibinfo{pages}{661} (\bibinfo{year}{1991}).

\bibitem{bennett:92}
\bibinfo{author}{C.~H. Bennett,}
\newblock \emph{\bibinfo{journal}{Phys. Rev. Lett.}}
  \textbf{\bibinfo{volume}{68}}, \bibinfo{pages}{3121} (\bibinfo{year}{1992}).

\bibitem{gisin:02}
\bibinfo{author}{N. Gisin}, \bibinfo{author}{G. Ribordy},
  \bibinfo{author}{W. Tittel} and \bibinfo{author}{H. Zbinden},
\newblock \emph{\bibinfo{journal}{Rev. Mod. Phys.}}
  \textbf{\bibinfo{volume}{74}}, \bibinfo{pages}{145} (\bibinfo{year}{2002}).

\bibitem{scarani:09}
\bibinfo{author}{V. Scarani}, \bibinfo{author}{H. Bechmann-Pasquinucci}, \bibinfo{author}{N.~J. Cerf}, \bibinfo{author}{M. Du\v{s}ek}, \bibinfo{author}{N. L\"utkenhaus} and \bibinfo{author}{M. Peev},
\newblock \emph{\bibinfo{journal}{Rev. Mod. Phys.}}
  \textbf{\bibinfo{volume}{81}}, \bibinfo{pages}{1301} (\bibinfo{year}{2009}).

\bibitem{muller:97}
\bibinfo{author}{A. Muller}, \bibinfo{author}{T. Herzog}, \bibinfo{author}{B. Huttner}, \bibinfo{author}{W. Tittel}, \bibinfo{author}{H. Zbinden} and \bibinfo{author}{N. Gisin},
\newblock \emph{\bibinfo{journal}{Appl. Phys. Lett.}}
  \textbf{\bibinfo{volume}{70}}, \bibinfo{pages}{793--795}
  (\bibinfo{year}{1997}).

\bibitem{scarani:04}
\bibinfo{author}{V. Scarani}, \bibinfo{author}{A. Acin},
  \bibinfo{author}{G. Ribordy} and \bibinfo{author}{N. Gisin},
  \newblock \emph{\bibinfo{journal}{Phys. Rev. Lett.}}
  \textbf{\bibinfo{volume}{92}}, \bibinfo{pages}{057901}
  (\bibinfo{year}{2004}).

\bibitem{brassard:00}
\bibinfo{author}{G. Brassard}, \bibinfo{author}{N. L{\"u}tkenhaus},
  \bibinfo{author}{T. Mor} and \bibinfo{author}{B.~C. Sanders},
\newblock \emph{\bibinfo{journal}{Phys. Rev. Lett.}}
  \textbf{\bibinfo{volume}{85}}, \bibinfo{pages}{1330} (\bibinfo{year}{2000}).

\bibitem{hwang:03}
\bibinfo{author}{W.-Y. Hwang},
\newblock \emph{\bibinfo{journal}{Phys. Rev. Lett.}}
  \textbf{\bibinfo{volume}{91}}, \bibinfo{pages}{057901}
  (\bibinfo{year}{2003}).

\bibitem{stucki:05}
\bibinfo{author}{D. Stucki}, \bibinfo{author}{N. Brunner},
  \bibinfo{author}{N. Gisin}, \bibinfo{author}{V. Scarani} and
\bibinfo{author}{H. Zbinden},
\newblock \emph{\bibinfo{journal}{Appl. Phys. Lett.}}
  \textbf{\bibinfo{volume}{87}}, \bibinfo{pages}{194108}
  (\bibinfo{year}{2005}).

\bibitem{bennett:95}
\bibinfo{author}{C.~H. Bennett}, \bibinfo{author}{G. Brassard},
  \bibinfo{author}{C. Cr{\'e}peau} and \bibinfo{author}{U.~M. Maurer},
\newblock \emph{\bibinfo{journal}{IEEE T. Inform. Theory}}
  \textbf{\bibinfo{volume}{41}}, \bibinfo{pages}{1915--1923}
  (\bibinfo{year}{1995}).

\bibitem{bechmann:00a}
\bibinfo{author}{H. Bechmann-Pasquinucci} and \bibinfo{author}{A. Peres},
\newblock \emph{\bibinfo{journal}{Phys. Rev. Lett.}}
  \textbf{\bibinfo{volume}{85}}, \bibinfo{pages}{3313} (\bibinfo{year}{2000}).

\bibitem{bechmann:00b}
\bibinfo{author}{H. Bechmann-Pasquinucci} and \bibinfo{author}{W. Tittel},
\newblock \emph{\bibinfo{journal}{Phys. Rev. A}} \textbf{\bibinfo{volume}{61}},
  \bibinfo{pages}{062308} (\bibinfo{year}{2000}).

\bibitem{cerf:02}
\bibinfo{author}{N.~J. Cerf}, \bibinfo{author}{M. Bourennane},
  \bibinfo{author}{A. Karlsson} and \bibinfo{author}{N. Gisin},
\newblock \emph{\bibinfo{journal}{Phys. Rev. Lett.}}
  \textbf{\bibinfo{volume}{88}}, \bibinfo{pages}{127902}
  (\bibinfo{year}{2002}).

\bibitem{inoue:02}
\bibinfo{author}{K. Inoue}, \bibinfo{author}{E. Waks} and
\bibinfo{author}{Y. Yamamoto},
\newblock \emph{\bibinfo{journal}{Phys. Rev. Lett.}}
  \textbf{\bibinfo{volume}{89}}, \bibinfo{pages}{037902}
  (\bibinfo{year}{2002}).

\bibitem{chau:15}
\bibinfo{author}{H.~F. Chau},
\newblock \emph{\bibinfo{journal}{Phys. Rev. A}} \textbf{\bibinfo{volume}{92}},
  \bibinfo{pages}{062324} (\bibinfo{year}{2015}).

\bibitem{chau:17}
\bibinfo{author}{H.~F. Chau}, \bibinfo{author}{Q. Wang} and \bibinfo{author}{C. Wong},
\newblock \emph{\bibinfo{journal}{Phys. Rev. A}} \textbf{\bibinfo{volume}{95}},
  \bibinfo{pages}{022311} (\bibinfo{year}{2017}).

\bibitem{wang:17}
\bibinfo{author}{S. Wang}, \bibinfo{author}{Z.-Q. Yin}, \bibinfo{author}{H.~F. Chau}, \bibinfo{author}{W. Chen}, \bibinfo{author}{C. Wang}, \bibinfo{author}{G.-C. Guo} and \bibinfo{author}{Z.-F. Han}, \newblock \eprint{arXiv:1707.00387}.

\bibitem{bacco:16}
\bibinfo{author}{D. Bacco}, \bibinfo{author}{J.~B. Christensen}, \bibinfo{author}{M.~A.~U. Castaneda}, \bibinfo{author}{Y. Ding}, \bibinfo{author}{S. Forchhammer}, \bibinfo{author}{K. Rottwitt} and \bibinfo{author}{L.~K. Oxenl{\o}we},
\newblock \emph{\bibinfo{journal}{Sci. Rep.}} \textbf{\bibinfo{volume}{6}},
  \bibinfo{pages}{36756} (\bibinfo{year}{2016}).

\bibitem{sasaki:14}
\bibinfo{author}{T. Sasaki}, \bibinfo{author}{Y. Yamamoto} and \bibinfo{author}{M. Koashi}, \newblock \emph{\bibinfo{journal}{Nature}} \textbf{\bibinfo{volume}{509}},
  \bibinfo{pages}{475} (\bibinfo{year}{2014}).

\bibitem{guan:15}
\bibinfo{author}{J.-Y. Guan}, \bibinfo{author}{Z. Cao}, \bibinfo{author}{Y. Liu}, \bibinfo{author}{G.-L. Shen-Tu}, \bibinfo{author}{J.~S. Pelc}, \bibinfo{author}{M.~M. Fejer}, \bibinfo{author}{C.-Z. Peng}, \bibinfo{author}{X. Ma}, \bibinfo{author}{Q. Zhang} and \bibinfo{author}{J.-W. Pan},
\newblock \emph{\bibinfo{journal}{Phys. Rev. Lett.}}
  \textbf{\bibinfo{volume}{114}}, \bibinfo{pages}{180502}
  (\bibinfo{year}{2015}).

\bibitem{takesue:15}
\bibinfo{author}{H. Takesue}, \bibinfo{author}{T. Sasaki},
  \bibinfo{author}{K. Tamaki} and \bibinfo{author}{M. Koashi},
\newblock \emph{\bibinfo{journal}{Nat. Photonics}}
  \textbf{\bibinfo{volume}{9}}, \bibinfo{pages}{827} (\bibinfo{year}{2015}).

\bibitem{wang:15}
\bibinfo{author}{S. Wang}, \bibinfo{author}{Z.-Q. Yin}, \bibinfo{author}{W. Chen}, \bibinfo{author}{D.-Y. He}, \bibinfo{author}{X.-T. Song}, \bibinfo{author}{H.-W. Li}, \bibinfo{author}{L.-J. Zhang}, \bibinfo{author}{Z. Zhou}, \bibinfo{author}{G.-C. Guo} and \bibinfo{author}{Z.-F. Han},
\newblock \emph{\bibinfo{journal}{Nat. Phys.}} \textbf{\bibinfo{volume}{9}},
  \bibinfo{pages}{832} (\bibinfo{year}{2015}).

\bibitem{li:16}
\bibinfo{author}{Y.-H. Li}, \bibinfo{author}{Y. Cao}, \bibinfo{author}{H. Dai}, \bibinfo{author}{J. Lin}, \bibinfo{author}{Z. Zhang}, \bibinfo{author}{W. Chen}, \bibinfo{author}{Y. Xu}, \bibinfo{author}{J.-Y. Guan}, \bibinfo{author}{S.-K. Liao}, \bibinfo{author}{J. Yin}, \bibinfo{author}{Q. Zhang}, \bibinfo{author}{X. Ma}, \bibinfo{author}{C.-Z. Peng} and \bibinfo{author}{J.-W. Pan},
\newblock \emph{\bibinfo{journal}{Phys. Rev. A}} \textbf{\bibinfo{volume}{93}},
  \bibinfo{pages}{030302} (\bibinfo{year}{2016}).

\bibitem{Yin:18}
\bibinfo{author}{Z.-Q. Yin}, \bibinfo{author}{S. Wang}, \bibinfo{author}{W. Chen}, \bibinfo{author}{Y.-G. Han}, \bibinfo{author}{R. Wang}, \bibinfo{author}{G.-C. Guo} and \bibinfo{author}{Z.-Fu Han},
\newblock \emph{\bibinfo{journal}{Nat. Commun.}} \textbf{\bibinfo{volume}{9}},
  \bibinfo{pages}{457} (\bibinfo{year}{2018}).

\bibitem{allen:92}
\bibinfo{author}{L. Allen}, \bibinfo{author}{M.~W. Beijersbergen},
  \bibinfo{author}{R. Spreeuw} and \bibinfo{author}{J. Woerdman},
\newblock \emph{\bibinfo{journal}{Phys. Rev. A}} \textbf{\bibinfo{volume}{45}},
  \bibinfo{pages}{8185} (\bibinfo{year}{1992}).

\bibitem{erhard:17}
\bibinfo{author}{M. Erhard}, \bibinfo{author}{R. Fickler},
  \bibinfo{author}{M. Krenn} and \bibinfo{author}{A. Zeilinger},
\newblock \emph{\bibinfo{journal}{Light: Sci. \& App.}}
  (\bibinfo{year}{2018}).

\bibitem{heckenberg:92}
\bibinfo{author}{N. Heckenberg}, \bibinfo{author}{R. McDuff},
  \bibinfo{author}{C. Smith} and \bibinfo{author}{A. White},
\newblock \emph{\bibinfo{journal}{Opt. Lett.}} \textbf{\bibinfo{volume}{17}},
  \bibinfo{pages}{221--223} (\bibinfo{year}{1992}).

\bibitem{forbes:16}
\bibinfo{author}{A. Forbes}, \bibinfo{author}{A. Dudley} and
\bibinfo{author}{M. McLaren},
\newblock \emph{\bibinfo{journal}{Advances in Optics and Photonics}}
  \textbf{\bibinfo{volume}{8}}, \bibinfo{pages}{200--227}
  (\bibinfo{year}{2016}).

\bibitem{mair:01}
\bibinfo{author}{A. Mair}, \bibinfo{author}{A. Vaziri},
  \bibinfo{author}{G. Weihs} and \bibinfo{author}{A. Zeilinger},
\newblock \emph{\bibinfo{journal}{Nature}} \textbf{\bibinfo{volume}{412}},
  \bibinfo{pages}{313--316} (\bibinfo{year}{2001}).

\bibitem{qassim:14}
\bibinfo{author}{H. Qassim}, \bibinfo{author}{F.~M. Miatto}, \bibinfo{author}{J.~P. Torres}, \bibinfo{author}{M.~J. Padgett}, \bibinfo{author}{E. Karimi} and \bibinfo{author}{R.~W. Boyd},
\newblock \emph{\bibinfo{journal}{J. Opt. Soc. Am. B}}
  \textbf{\bibinfo{volume}{31}}, \bibinfo{pages}{A20--A23}
  (\bibinfo{year}{2014}).

\bibitem{berkhout:10}
\bibinfo{author}{G.~C. Berkhout}, \bibinfo{author}{M.~P. Lavery}, \bibinfo{author}{J. Courtial}, \bibinfo{author}{M.~W. Beijersbergen} and \bibinfo{author}{M.~J. Padgett},
\newblock \emph{\bibinfo{journal}{Phys. Rev. Lett.}}
  \textbf{\bibinfo{volume}{105}}, \bibinfo{pages}{153601}
  (\bibinfo{year}{2010}).

\bibitem{larocque:17}
\bibinfo{author}{H. Larocque}, \bibinfo{author}{J. Gagnon-Bischoff}, \bibinfo{author}{D. Mortimer}, \bibinfo{author}{Y. Zhang}, \bibinfo{author}{F. Bouchard}, \bibinfo{author}{J. Upham}, \bibinfo{author}{V. Grillo}, \bibinfo{author}{R.~W. Boyd} and \bibinfo{author}{E. Karimi},
\newblock \emph{\bibinfo{journal}{Optics Express}}
  \textbf{\bibinfo{volume}{25}}, \bibinfo{pages}{19832--19843}
  (\bibinfo{year}{2017}).

\bibitem{mafu:13}
\bibinfo{author}{M. Mafu},\bibinfo{author}{A. Dudley}, \bibinfo{author}{S. Goyal}, \bibinfo{author}{D. Giovannini}, \bibinfo{author}{M. McLaren}, \bibinfo{author}{M.~J Padgett}, \bibinfo{author}{T. Konrad}, \bibinfo{author}{F. Petruccione}, \bibinfo{author}{N. L\"utkenhaus}, \bibinfo{author}{A. Forbes},
\newblock \emph{\bibinfo{journal}{Phys. Rev. A}} \textbf{\bibinfo{volume}{88}},
  \bibinfo{pages}{032305} (\bibinfo{year}{2013}).

\bibitem{mirhosseini:15}
\bibinfo{author}{M. Mirhosseini}, \bibinfo{author}{O.~S. Maga{\~n}a-Loaiza}, \bibinfo{author}{M.~N. O'Sullivan}, \bibinfo{author}{B. Rodenburg}, \bibinfo{author}{M. Malik}, \bibinfo{author}{M.~P.~J. Lavery}, \bibinfo{author}{M.~J. Padgett}, \bibinfo{author}{D.~J. Gauthier}, \bibinfo{author}{R.~W. Boyd},\newblock \emph{\bibinfo{journal}{New J. Phys.}} \textbf{\bibinfo{volume}{17}},
  \bibinfo{pages}{033033} (\bibinfo{year}{2015}).

\bibitem{vallone:14}
\bibinfo{author}{G. Vallone}, \bibinfo{author}{V. D'Ambrosio}, \bibinfo{author}{A. Sponselli}, \bibinfo{author}{S. Slussarenko}, \bibinfo{author}{L. Marrucci}, \bibinfo{author}{F. Sciarrino} and \bibinfo{author}{P. Villoresi},
\newblock \emph{\bibinfo{journal}{Phys. Rev. Lett.}}
  \textbf{\bibinfo{volume}{113}}, \bibinfo{pages}{060503}
  (\bibinfo{year}{2014}).

\bibitem{krenn:15}
\bibinfo{author}{M. Krenn}, \bibinfo{author}{J. Handsteiner},
\bibinfo{author}{M. Fink}, \bibinfo{author}{R. Fickler} and \bibinfo{author}{A. Zeilinger},
\newblock \emph{\bibinfo{journal}{PNAS}} \textbf{\bibinfo{volume}{112}},
  \bibinfo{pages}{14197--14201} (\bibinfo{year}{2015}).

\bibitem{sit:17}
\bibinfo{author}{A. Sit}, \bibinfo{author}{F. Bouchard}, \bibinfo{author}{R. Fickler}, \bibinfo{author}{J. Gagnon-Bischoff}, \bibinfo{author}{H. Larocque}, \bibinfo{author}{K. Heshami}, \bibinfo{author}{D. Elser}, \bibinfo{author}{C. Peuntinger}, \bibinfo{author}{K. G\"unthner}, \bibinfo{author}{B. Heim}, \bibinfo{author}{C. Marquardt}, \bibinfo{author}{G. Leuchs}, \bibinfo{author}{R.~W. Boyd} and \bibinfo{author}{E. Karimi},
\newblock \emph{\bibinfo{journal}{Optica}} \textbf{\bibinfo{volume}{4}},
  \bibinfo{pages}{1006--1010} (\bibinfo{year}{2017}).

\bibitem{bouchard:18}
\bibinfo{author}{F. Bouchard}, \bibinfo{author}{A. Sit}, \bibinfo{author}{F. Hufnagel}, \bibinfo{author}{A. Abbas}, \bibinfo{author}{Y. Zhang}, \bibinfo{author}{K. Heshami}, \bibinfo{author}{R. Fickler}, \bibinfo{author}{C. Marquardt}, \bibinfo{author}{G. Leuchs}, \bibinfo{author}{R.~W. Boyd} and \bibinfo{author}{E. Karimi}, \newblock \eprint{arXiv:1801.10299}.

\bibitem{born:13}
\bibinfo{author}{M. Born} and \bibinfo{author}{E. Wolf}, (\bibinfo{publisher}{Elsevier}, \bibinfo{year}{2013}).

\bibitem{karimi:14b}
\bibinfo{author}{E. Karimi}, \bibinfo{author}{D. Giovannini}, \bibinfo{author}{E. Bolduc}, \bibinfo{author}{N. Bent}, \bibinfo{author}{F.~M. Miatto}, \bibinfo{author}{M.~J. Padgett} and \bibinfo{author}{R.~W. Boyd},
\newblock \emph{\bibinfo{journal}{Physical Review A}}
  \textbf{\bibinfo{volume}{89}}, \bibinfo{pages}{013829}
  (\bibinfo{year}{2014}).

\bibitem{karimi:14c} 
\bibinfo{author}{E. Karimi}, \bibinfo{author}{R.~W. Boyd}, \bibinfo{author}{P. de la Hoz}, \bibinfo{author}{H. de Guise}, \bibinfo{author}{J. \v{R}eh\'a\v{c}ek} \bibinfo{author}{Z. Hradil}, \bibinfo{author}{A. Aiello}, \bibinfo{author}{G. Leuchs} and \bibinfo{author}{L. L. S\`anchez-Soto},
\newblock \emph{\bibinfo{journal}{Physical review A}}
  \textbf{\bibinfo{volume}{89}}, \bibinfo{pages}{063813}
  (\bibinfo{year}{2014}).

\bibitem{marrucci:06}
\bibinfo{author}{L. Marrucci}, \bibinfo{author}{C. Manzo} and \bibinfo{author}{D. Paparo},
\newblock \emph{\bibinfo{journal}{Phys. Rev. Lett.}}
  \textbf{\bibinfo{volume}{96}}, \bibinfo{pages}{163905}
  (\bibinfo{year}{2006}).

\bibitem{mirhosseini:13b}
\bibinfo{author}{M. Mirhosseini}, \bibinfo{author}{O.~S Magana-Loaiza}, \bibinfo{author}{C. Chen}, \bibinfo{author}{B. Rodenburg}, \bibinfo{author}{M. Malik}, \bibinfo{author}{R.~W. Boyd},
\newblock \emph{\bibinfo{journal}{Opt. Express}} \textbf{\bibinfo{volume}{21}},
  \bibinfo{pages}{30196--30203} (\bibinfo{year}{2013}).

\bibitem{beijersbergen:94}
\bibinfo{author}{M. Beijersbergen}, \bibinfo{author}{R. Coerwinkel}, \bibinfo{author}{M. Kristensen} and \bibinfo{author}{J. Woerdman},
\newblock \emph{\bibinfo{journal}{Opt. Commun.}}
  \textbf{\bibinfo{volume}{112}}, \bibinfo{pages}{321--327}
  (\bibinfo{year}{1994}).

\bibitem{yu:11}
\bibinfo{author}{N. Yu}, \bibinfo{author}{P. Genevet}, \bibinfo{author}{M.~A. Kats}, \bibinfo{author}{F. Aieta}, \bibinfo{author}{J.-P. Tetienne}, \bibinfo{author}{F. Capasso} and \bibinfo{author}{Z. Gaburro},
\newblock \emph{\bibinfo{journal}{Science}} \textbf{\bibinfo{volume}{334}},
  \bibinfo{pages}{333--337} (\bibinfo{year}{2011}).

\bibitem{karimi:14}
\bibinfo{author}{E. Karimi}, \bibinfo{author}{S.~A. Schulz}, \bibinfo{author}{I.~D. Leon}, \bibinfo{author}{H. Qassim}, \bibinfo{author}{J. Upham}, \bibinfo{author}{R.~W. Boyd},
\newblock \emph{\bibinfo{journal}{Light: Sci. \& App.}}
  \textbf{\bibinfo{volume}{3}}, \bibinfo{pages}{e167} (\bibinfo{year}{2014}).

\bibitem{bouchard:14}
\bibinfo{author}{F. Bouchard}, \bibinfo{author}{I.~D. Leon}, \bibinfo{author}{S.~A. Schulz}, \bibinfo{author}{J. Upham}, \bibinfo{author}{E. Karimi}, \bibinfo{author}{R.~W. Boyd},
\newblock \emph{\bibinfo{journal}{Appl. Phys. Lett.}}
  \textbf{\bibinfo{volume}{105}}, \bibinfo{pages}{101905}
  (\bibinfo{year}{2014}).

\bibitem{durt:10}
\bibinfo{author}{T. Durt}, \bibinfo{author}{B.-G. Englert},
  \bibinfo{author}{I. Bengtsson} and \bibinfo{author}{K. {\.Z}yczkowski},
\newblock \emph{\bibinfo{journal}{Int. J. Quantum Inf.}}
  \textbf{\bibinfo{volume}{8}}, \bibinfo{pages}{535--640}
  (\bibinfo{year}{2010}).

\bibitem{englert:04b}
\bibinfo{author}{B.-G. Englert}, \bibinfo{author}{D. Kaszlikowski}, \bibinfo{author}{H.~K. Ng}, \bibinfo{author}{W.~K. Chua}, \bibinfo{author}{J. \v{R}eh\'a\v{c}ek}, \bibinfo{author}{J. Anders}, \newblock \eprint{arXiv:0412075}.

\bibitem{bouchard:18b}
\bibinfo{author}{F. Bouchard}, \bibinfo{author}{K. Heshami}, \bibinfo{author}{D. England}, \bibinfo{author}{R. Fickler}, \bibinfo{author}{R.~W. Boyd}, \bibinfo{author}{B.-G. Englert}, \bibinfo{author}{L.~L. S\'anchez-Soto}, \bibinfo{author}{E. Karimi}, \newblock \eprint{arXiv:1802.05773}.

\end{thebibliography}

\end{document}